\documentstyle[11pt,aaspp4]{article}

\def\chi{\mu_e}

\begin{document}
\title{Emission from Bow Shocks of Beamed Gamma-Ray Bursts}
\author{Xiaohu Wang and Abraham Loeb}
\medskip
\affil{Harvard-Smithsonian Center for Astrophysics, 60 Garden Street,
Cambridge, MA 02138; xwang@cfa.harvard.edu, aloeb@cfa.harvard.edu}

\begin{abstract}

Beamed $\gamma$-ray burst (GRB) sources produce a bow shock in their
gaseous environment. The emitted flux from this bow shock may dominate over
the direct emission from the jet for lines of sight which are outside the
angular radius of the jet emission, $\Theta_b$.  The event rate for these
lines of sight is increased by a factor of $\sim 260
(\Theta_b/5^\circ)^{-2}$.  For typical GRB parameters, we find that the bow
shock emission from a jet of half-angle $\sim 5^\circ$ is visible out to
tens of Mpc in the radio and hundreds of Mpc in the X-rays. If GRBs are
linked to supernovae, studies of peculiar supernovae in the local universe
should reveal this non-thermal bow shock emission for weeks to months
following the explosion.

\end{abstract} 


\section{INTRODUCTION}

The afterglows of Gamma-ray bursts (GRBs) are most naturally described by
the so-called ``fireball'' model (see e.g., Paczy\'{n}ski \& Rhoads 1993;
Katz 1994; M\'{e}sz\'{a}ros \& Rees 1993, 1997; Waxman 1997a,b; Sari,
Piran, \& Narayan 1998). In this model, a compact source releases a large
amount of energy over a short time and produces a relativistically
expanding fireball. Eventually the fireball interacts with the circumburst
medium, producing a spherical relativistic shock in it.  As the shock
decelerates due to the accumulation of mass from the external medium, it
approaches a self-similar solution (Blandford \& McKee 1976) and produces
delayed synchrotron emission in X-rays, optical and radio, similar to the
observed afterglows.  The model generically yields power-law spectra and
lightcurves in general agreement with observations at X-ray (Costa et
al.~1997), optical (van Paradijs et al. 1997) and radio (Frail et al.~1997)
wavelengths.  Precise positions have allowed redshifts to be measured for a
number of GRBs (Metzger et al.~1997), providing a definitive proof of their
cosmological origin.

Recent observational evidence indicates that at least some GRB are not
spherical explosions.  A broad-band break in the lightcurve power-law index
was predicted for shocks produced by collimated jets due to the lateral
expansion of the jet when its Lorentz factor drops below the inverse of its
opening angle (Rhoads 1997, 1999a,b; Panaitescu \& M\'{e}sz\'{a}ros 1999). 
Such breaks have been seen in GRB~990510 (Stanek et al. 1999; Harrison et
al. 1999), GRB~991216 (Halpern et al. 2000), and GRB~000301C (Sagar et
al. 2000; Masetti et al. 2000; Jensen et al. 2000; Berger et al. 2000).
The ratio of the spectral index to the light curve index also suggests
non-spherical energy ejection for some events (e.g. GRB~991216: Garnavich
et al. 2000).

The possibility that GRBs are associated with a rare sub-class of
supernovae (SNe) was advocated recently based on a possible SN component in
the lightcurves of GRB 980326 (Bloom et al. 1999) and GRB 970228 (Reichart
1999) and the association of SN1998bw with GRB 980425 (Galama et al. 1999;
Kulkarni et al. 2000).  In this case the GRB emission needs to be beamed in
order to achieve the high Lorentz factor required for GRB outflows, despite
the massive progenitor envelope and the low energy yield of SNe (MacFadyen,
Woosley, \& Heger 1999; Khokhlov at al. 1999; Umeda 2000).

The current state of knowledge leaves two major questions open: (i) are
GRBs associated with rare SNe?; and (ii) if so, what is the characteristic
collimation angle of the energy release? The collimation angle has
important implications on the nature of the central engine.  If the average
angular radius of the two opposing jets in collimated GRBs is $\Theta_b$,
then the total energy release will be reduced by a factor $2
\pi\Theta_b^2/4\pi=\Theta_b^2/2$ and the event rate will be increased by
the inverse of this factor. In this paper we propose a direct observational
probe that will help answer these questions.

When a transient relativistic jet (equivalent to a relativistic ``bullet'')
moves into the circumburst medium (be it the interstellar medium or a
progenitor wind), a bow shock is generated in front of the jet.  The bow
shock accelerates electrons which emit synchrotron radiation, similarly to
the primary shock in front of the jet. For observers situated within the
cone of the jet emission, the bow shock emission will be sub-dominant
relative to the jet emission due to relativistic motion of the jet towards
the observer.  However, for highly collimated outflows, most lines of sight
lie outside the jet cone, and for those -- the bow shock emission may
dominate.  In particular, if GRBs originate in SNe then the bow shock
emission will add a strong non-thermal component to the emission from the
SN remnants extending from the radio to the X-ray bands.  This emission
will extend over much longer times compared to the jet emission since the
Lorentz factor of the bow shock is typically much smaller than that of the
jet, resulting in a smaller relativistic compression of time in the
observer's frame.  In the following sections we will apply the same model
used ordinarily for the primary fireball shock, to calculate the flux in
this non-thermal bow-shock component.  In \S 2 we present our model for the
bow shock emission; in \S 3 we show our numerical results; and finally, in
\S 4, we summarize our main conclusions.

\section{MODEL} 

The impulsive ejection of a relativistic jet from the GRB source results in
a thin disk of material moving outwards and expanding, which we refer to as
a {\it relativistic bullet}.  As long as the angular radius of the bullet
relative to the center of the explosion, $\Theta_b$, is much larger than
the inverse of its Lorentz factor, the bullet will expand as if it is part
of a spherically symmetric fireball. This follows from the relativistic
transformation of velocities, which implies that the propagation speed of a
signal in the perpendicular direction to the bullet motion cannot exceed
$\sim c/\gamma_b$ in the frame of the stationary ambient medium. For
simplicity, we limit our discussion to this regime.

In our model we assume that as the relativistic {\it bullet} moves through
the ambient medium, it deposits a fraction of its kinetic energy in each
infinitesimal distance it travels. We treat this energy deposition as a
sequence of point explosions, which drive shocks into the surrounding
medium. The combination of this sequence of small shocks results in a
conical bow shock structure behind the bullet. 

Figure 1 illustrates the geometry of the bow shock. We use cylindrical
coordinates, with the $z$-axis connecting the bullet center to the
source. The distance between the source and the bullet is denoted as
$R_b$. We assume that observers are located perpendicular to the $z$-axis,
with $\theta = 0$ along the source-observer axis.  Next we consider a slice
of the bow shock geometry perpendicular to the $z$-axis. The slice is a
cylindrical shell of height $\delta z$, thickness $\Delta$ and inner radius
$r$. As the bullet passes through a point $z$ it has a radius $l(z)$ and a
Lorentz factor $\gamma_b(z)$. The amount of thermal energy $\delta E$ which
gets deposited by the bullet in the surrounding medium as it moves an extra
distance $\delta z$, is then given by
\begin{equation}
\delta E = (\gamma_b-1)\pi l^2\ \delta z\ n\chi m_pc^2 ,
\label{eq:dE}
\end{equation}
where $n$ is the electron number density of the external medium, and $\chi
m_p$ is the ion mass per electron. We assume that a fraction $\varepsilon$
of the energy stored in the ``causally-connected'' edge of the bullet is
transferred to the external medium. The ``causally-connected'' edge of the
bullet is defined to be the region occupying an angular 
size $1/\gamma_b$ from the bullet edge.  The 
energy deposited in the infinitesimal distance
interval $\delta z$ drives a shock into the ambient gas. The gas behind the
shock is piled into an outgoing thin shell. The shock expands perpendicular
to the $z$-axis, but due to the time delay between the energy deposition at
different points along the $z$-axis, the sum of the fronts generated by
these points defines a conical shape for the resulting bow shock.  We
denote the Lorentz factor of the outgoing shell in the frame of the
unshocked gas by $\gamma$ . The electron number density $n'$ and internal
energy density $e'$ of the shocked gas in the frame comoving with the shell
can be written as (Blandford \& McKee 1976)
\begin{equation}
n' = \frac{\hat{\gamma}\gamma+1}{\hat{\gamma}-1}n ,
\label{eq:n}
\end{equation}
\begin{equation}
e' = \frac{\hat{\gamma}\gamma+1}{\hat{\gamma}-1}(\gamma-1)n\chi m_pc^2 ,
\label{eq:e}
\end{equation}
where $\hat{\gamma}$ is the adiabatic index of the shocked gas which
changes from $4/3$ and $5/3$ as the shock velocity changes from the
relativistic to the non-relativistic regime. We interpolate between these
regimes using the simplified expression $\hat{\gamma} \approx
(4\gamma+1)/(3\gamma)$ (Dai, Huang \& Lu 1999). Equations (\ref{eq:n}) and
(\ref{eq:e}) then become
\begin{equation}
n' = 4\gamma n ,
\label{eq:nnew}
\end{equation}
\begin{equation}
e' = 4b\gamma^2n\chi m_pc^2 ,
\label{eq:enew}
\end{equation}
where $b = (\gamma-1)/\gamma$. Note that equations (\ref{eq:nnew}) and
(\ref{eq:enew}) are appropriate for both the relativistic and
non-relativistic regimes. The total kinetic energy of the shell is
\begin{equation}
E_k = \pi (r^2-l^2)\ \delta z\ n\chi m_pc^2\beta^2\gamma^2 ,
\label{eq:Ek}
\end{equation}
where $\beta^2 = 1-1/\gamma^2$. Conservation of energy implies, $E_k =
\varepsilon \xi \delta E$, where
\begin{equation}
\xi = \frac{2\Theta_b \gamma_b - 1}{\Theta_b^2 \gamma_b^2}
\ \ \ \ {\rm for} \ \gamma_b \ge \frac{1}{\Theta_b}
\label{eq:xi}
\end{equation}
is the ratio between the volumes of the causaly-connected region near the
edge of the bullet (with angular size $1/\gamma_b$) and the entire volume
of the bullet.
We therefore obtain
\begin{equation}
\gamma = \sqrt{1+ \varepsilon \xi (\gamma_b-1)\frac{l^2}{r^2-l^2}} =
\sqrt{1+ f\frac{l^2}{r^2-l^2}} ,
\label{eq:gamma}
\end{equation}
where $f=\varepsilon \xi (\gamma_b-1)$. The thickness of the shell is
derived by using the conservation of particle number. This yields
\begin{equation}
\Delta = \frac{r^2-l^2}{8\gamma^2r}.
\label{eq:Delta}
\end{equation}

Figure 2 illustrates the geometry of the emission from an infinitesimal
volume element in cylindrical coordinates, $dV=rdr d\theta dz$.  We define
the emission coefficient $j'_{\nu'}$ to be the power emitted per unit
frequency, $\nu'$, per unit volume per steradian in the rest frame of the
emitting material.  We use prime to denote quantities in the local rest
frame of the emitting material, while unprimed quantities are measured in
the rest frame of the external medium.  Note that $j_{\nu} / \nu^{2}$ is
Lorentz invariant (Rybicki \& Lightman 1979).  For each slice, the
cylindrical shell of expanding material emits isotropically in its local
rest frame with $j'_{\nu'}=P'({\nu'},r,t)/4\pi$ and $\nu'=\nu \gamma
(1-\beta \mu)$, where $\gamma$ and $\beta c$ are the Lorentz factor and the
velocity of the emitting material, and $\mu=\cos \theta$. A photon emitted
at time $t$ and place $\bf{\sl r}$ in the unshocked gas frame will reach
the detector at a time $T$ given by
\begin{equation}
T_Z=\frac{T}{1+Z}=t-\frac{r \mu}{c},
\label{eq:time}
\end{equation}
where $Z$ is the cosmological redshift of the GRB and $T$ is chosen so that
a photon emitted at the GRB source at $t=0$ will arrive to the detector at
$T=0$. Thus we have \footnote{Note that a similar equation for calculating
the emission from GRB afterglows in a spherical coordinate system was
originally derived by Granot, Piran, \& Sari (1999).}
\begin{equation}
F(\nu, T)=2 \times \frac{1+Z}{4\pi D^{2}}\int_{0}^{\infty} dz 
\int_{-\pi}^{\pi} d\theta \int_{0}^{\infty} rdr 
\frac{P'[\nu \gamma (1-\beta \mu), r, T_{Z}+{r\mu}/{c}]}
{\gamma ^{2} (1-\beta \mu)^{2}},
\label{eq:flux}
\end{equation}
where $D$ is the luminosity distance to the GRB, and $\gamma, \beta,
\mu$ are evaluated at the time $t$ implied by equation~(\ref{eq:time}). The
factor of 2 is added to describe the combined emission from the bow shocks
generated by two opposing jets.

The volume integration expressed in equation~(\ref{eq:flux}) should be
taken over the region occupied by the emitting bow shock at a given
observed time.
The lower and upper boundaries of the integration over $r$ are derived by
considering the relativistic time delay (similar to the approach used by
Wang \& Loeb, 2000). For a given observed time $T$, the outer boundary
$R_1$ satisfies the equation
\begin{equation}
T = \frac{1}{c}F(R_1)-\frac{R_1\mu}{c} ,
\label{eq:outer}
\end{equation}
where 
\begin{equation}
F(R)=\frac{1}{\sqrt{f}l}\left\{\frac{1}{2}R
\sqrt{R^2+(f-1)l^2}-\frac{1}{2}\sqrt{f}l^2
+\frac{1}{2}(f-1)l^2\log\frac{R+\sqrt{R^2+(f-1)l^2}}{(\sqrt{f}+1)l}
\right\} .
\label{eq:FR}
\end{equation}
The inner boundary $R_2$ can be obtained by solving the equation
\begin{equation}
T = \frac{1}{c}F(R_2) -
\left\{R_2-\frac{(R_2^2-l^2)^2}{8R_2[R_2^2+(f-1)l^2]} 
\right\}\frac{\mu}{c}.
\label{eq:inner}
\end{equation}

Next we derive the local synchrotron emissivity in the two regimes where
the infinitesimal segment of the bow shock under consideration is moving at
relativistic or nonrelativistic speeds. In the relativistic regime, we
assume that the energy densities of the shock-accelerated electrons and the
magnetic field are fixed fractions of the internal energy density in the
shell behind the bow shock, $e'_e=\epsilon_ee',\ e'_B=\epsilon_Be'$, and
that the bow shock produces a power law distribution of accelerated
electrons with a number density per Lorentz factor of
$N(\gamma_e)=K\gamma_e^{-p}$ for $\gamma_e\geq\gamma_{\rm min}$, where
\begin{equation}
\gamma_{\rm min}=\left( \frac{p-2}{p-1} \right) 
\frac{\epsilon_ee'}{n'm_ec^2} , \ \ \ \ K=(p-1)n'\gamma_{\rm min}^{p-1}.
\label{eq:K1}
\end{equation}
When the bow shock is nonrelativistic, we use $\epsilon_e$ to denote the
fraction of the internal energy which goes to relativistic electrons with
$\gamma\ga 2$. Again, these relativistic electrons obtain a power law
distribution $N_{\rm rel}(\gamma_e)=K\gamma_e^{-p}$ for
$\gamma_e\geq\gamma_{\rm min}$, where
\begin{equation}
K = \epsilon_e \frac{m_p}{m_e} \chi (p-2) 4\gamma(\gamma-1)n\gamma_{\rm
min}^{p-2},
\label{eq:K2}
\end{equation}
and $\gamma_{\rm min}=2$. Hence the number density of relativistic
electrons is
\begin{equation}
n'_{\rm rel} = \epsilon_e \frac{m_p}{m_e} \chi \frac{p-2}{p-1}
4\gamma(\gamma-1)\frac{n}{\gamma_{\rm min}} .
\label{eq:nrel}
\end{equation}
This prescription describes well the accelerated electron population in the
non-relativistic shocks of supernovae (Chevalier 1999). The characteristic
values inferred for the parameters $p\sim 2$--$3$ and $\epsilon_e\sim
1$--$10\%$ in supernova shocks (Chevalier 1999; Koyama et al. 1995, 1997;
Tanimori et al. 1998; Muraishi et al. 2000) are in the same range as those
for GRB afterglows (Waxman 1997a,b; Wijers, Rees, \& Meszaros 1997; Sari,
Piran, \& Narayan al. 1998). We will consider different parameter values,
but for simplicity, we will adopt the same values of $\epsilon_e$ and $p$
in describing both the relativistic and sub-relativistic regimes.
This is justified by the similarity between the parameter values 
which are typically  chosen
to describe relativistic GRB afterglows (e.g., Waxman 1997a,b) 
and non-relativistic radio supernovae (e.g., Chevalier 1998).

At sufficiently high Lorentz factors, the electron cooling time is shorter
than the dynamical time. The critical Lorentz factor $\gamma_c$ above which
electrons cool on a time shorter than $t_{\rm dyn}$ is given by (Sari, 
Piran \& Narayan 1998)
\begin{equation}
\gamma_c = \frac{3m_e}{16\epsilon_B \sigma_T\chi m_pc} 
\frac{1}{t_{\rm dyn}b\gamma^3n}
\label{eq:gammacool}
\end{equation}
where $t_{\rm dyn}$ is equal to the dynamical time in the frame
of the external medium, i.e., the time it takes for
a cylindrical shell to expand from an initial radius $l$ to $r$,
\begin{equation}
t_{\rm dyn} = \frac{1}{c}F(r) .
\label{eq:t_c}
\end{equation}
Note that
equation~(\ref{eq:gammacool}) is different from equation (6) in Sari et al.
(1998) because of the introduction of $b$ and $\chi$.  An electron with an
initial Lorentz factor $\gamma_e > \gamma_c$ cools down to $\gamma_c$ in
the time $t_{\rm dyn}$.

The radiation power and the characteristic synchrotron frequency of a
randomly oriented electron with a Lorentz factor $\gamma_e$ are given by
(Rybicki \& Lightman 1979)
\begin{equation}
P'(\gamma_e) = \frac{4}{3} \sigma_T \gamma_e^2 \beta^2 c \frac{B'^2}{8\pi} ,
\label{eq:Pgammae}
\end{equation}
\begin{equation}
\nu'_{\rm syn}(\gamma_e) = \frac{3\gamma_e^2 q_e B'}{16 m_e c} ,
\label{eq:nusyn}
\end{equation}
where $B' = \sqrt{8\pi e'_B}$, and $m_e,q_e$ are the electron mass and
charge respectively.  
We define $\nu'_c=\nu'_{\rm syn}(\gamma_c)$ and $\nu'_m=\nu'_{\rm
syn}(\gamma_{\rm min})$. In the regime of fast cooling, $\gamma_{\rm min} >
\gamma_c$, the emissivity is given by (Sari et al. 1998)
\begin{equation}
P'_{\nu'} = \left\{ \begin{array}{ll}
P'_{\nu',{\rm max}}\left({\nu'}/{\nu'_c}\right)^{1/3} 
& \nu'<\nu'_c \\
P'_{\nu',{\rm max}}\left({\nu'}/{\nu'_c}\right)^{-1/2} 
& \nu'_c\le\nu'<\nu'_m \\
P'_{\nu',{\rm max}}\left({\nu'_m}/{\nu'_c}\right)^{-1/2}
 \left({\nu'}/{\nu'_m}\right)^{-p/2}
& \nu'\ge\nu'_m ,
\end{array}
\right.
\label{eq:fastcool}
\end{equation}
where 
\begin{equation}
P'_{\nu',{\rm max}} \approx n'\frac{P'(\gamma_e)}{\nu'_{\rm syn}(\gamma_e)} 
= \frac{8m_ec^2\sigma_T}{9\pi q_e}\beta^2n'B' .
\label{eq:pmax}
\end{equation}
In the above equation, $n'$ should be replaced by $n'_{\rm rel}$ for the
nonrelativistic regime. 

For slow cooling, $\gamma_c > \gamma_{\rm min}$, the emissivity is given by
\begin{equation}
P'_{\nu'} = \left\{ \begin{array}{ll}
P'_{\nu',{\rm max}}\left({\nu'}/{\nu'_m}\right)^{1/3} 
& \nu'<\nu'_m \\
P'_{\nu',{\rm max}}\left({\nu'}/{\nu'_m}\right)^{-(p-1)/2} 
& \nu'_m\le\nu'<\nu'_c \\
P'_{\nu',{\rm max}}\left({\nu'_c}/{\nu'_m}\right)^{-(p-1)/2}
 \left({\nu'}/{\nu'_c}\right)^{-p/2}
& \nu'\ge\nu'_c.
\end{array}
\right.
\label{eq:slowcool}
\end{equation}

At very low frequencies, synchrotron self-absorption becomes important.  In
the comoving frame of the shocked gas, the absorption coefficient
$\alpha'_{\nu'}$ scales as $\alpha'_{\nu'}\propto \nu'^{-(p+4)/2}$ for
$\nu' > \nu'_m$ and as $\alpha'_{\nu'}\propto \nu'^{-5/3}$ for $\nu' <
\nu'_m$ (Waxman 1997a). We therefore write
\begin{equation}
\alpha'_{\nu'} = H \nu'^{-(p+4)/2} ,\ \ \nu' > \nu'_m ,
\end{equation}
where (Rybicki \& Lightman 1979)
\begin{equation}
H\equiv \frac{\sqrt{3}q_e^3}{8\pi m_e} \left( \frac{3q_e}{2\pi
m_e^3 c^5} \right)^{p/2} (m_e c^2)^{p-1}K \lambda B'^{(p+2)/2}\Gamma
\left(\frac{3p+2}{12}\right) \Gamma \left(\frac{3p+22}{12}\right)
\label{eq:alphanu1}
\end{equation}
and where $\lambda = (1/2)\int^{\pi}_{0} (\sin\alpha)^{(p+2)/2}\sin \alpha
d\alpha$, and $\Gamma(y)$ is the Gamma function.  For $\nu' < \nu'_m$ we
then use
\begin{equation}
\alpha'_{\nu'} = H \nu'^{-(p+4)/2}_m \left( 
\frac{\nu'}{\nu'_m} \right)^{-5/3}, \ \ \ \ \ \nu' < \nu'_m.
\label{eq:alphanu2}
\end{equation}
Because $\nu \alpha_{\nu}$ is Lorentz invariant, the absorption coefficient
in the rest frame of the unshocked gas is
\begin{equation}
\alpha_{\nu} = \gamma(1-\beta \mu) \alpha'_{\nu'}.
\label{eq:alphanu3}
\end{equation}
Equation (\ref{eq:flux}) should then be modified to
\begin{equation}
F(\nu, T)=2\times \frac{1+Z}{4\pi D^{2}}\int_{0}^{\infty} dz 
\int_{-\pi}^{\pi} d\theta \int_{0}^{\infty} rdr 
\left(\frac{1-e^{-\tau_{\nu}}}{\tau_{\nu}} \right) 
\frac{P'[\nu', r, t]}{\gamma ^{2} (1-\beta \mu)^{2}},
\label{eq:fluxabsorption}
\end{equation}
where $\tau_{\nu} \approx \alpha_{\nu} \times \Delta$ is the optical depth
per unit frequency for synchrotron self-absorption across the shell
thickness.

The external medium could be either the interstellar medium (ISM) of the
host galaxy (Waxman 1997a,b) or a precurser wind that was ejected by the
GRB progenitor (Chevalier \& Li 1999; 2000). For a wind profile $\rho =
Ar^{-2}$, the distance between the source and the bullet is given by
(Chevalier \& Li 2000)
\begin{equation}
R_b = 1.1\times10^{17}\left(\frac{5.9}{\gamma_b}\right)^{2}E_{52}A_*^{-1}\
{\rm cm} ,
\label{eq:RB}
\end{equation}
where $\gamma_b$ is the bullet Lorentz factor, $E_{52}$ is the equivalent
isotropic energy release of the GRB in units of $10^{52}\ {\rm ergs}$,
$A=\dot{M}_{\rm w}/4\pi V_{\rm w} = 5\times10^{11}A_*\ {\rm g\ cm}^{-1}$,
$\dot{M}_{\rm w}$ is the progenitor mass loss rate, and $V_{\rm w}$ is the
wind velocity. $A_*=1$ corresponds to $\dot{M}_{\rm w} = 10^{-5}\
M_{\odot}\ {\rm yr}^{-1}$ and $V_{\rm w}=1000\ {\rm km}\ {\rm s}^{-1}$. The
electron number density is given by, $n = \rho / (\chi m_p)$. Also note
that the energy input into the two jets is equal to $E\times (\Theta_b^2 /
2)$.

If the external medium is the ISM, $R_b$ is given by (Granot, Piran, \&
Sari 1999)
\begin{equation}
R_b = 5.53\times10^{17}\left(\frac{3.65}{\gamma_b}\right)^{2/3}
\left(\frac{E_{52}}{n_1}\right)^{1/3}\ {\rm cm} ,
\label{eq:RBISM}
\end{equation}
where $n_1$ is the ISM electron number density in units of $1\ {\rm cm}^{-3}$.

\section{NUMERICAL RESULTS}

We have solved the equations of \S 2 for different values of the free
parameters in our model.  In the following, we show results for a standard
case, and five variations on the values of its parameters.  In the standard
case 1, we adopt the parameter values, $\Theta_b=0.1$, $E_{52}=1$,
$\varepsilon=0.1$, $\epsilon_e=0.1$, $\epsilon_B=0.1$. We also adopt
$A_*=1$ and $\chi=2$ for the wind of a typical Wolf-Rayet star (Chevalier 
\& Li 2000). Cases 2-5 also refer to a wind profile, but with a
variation on one parameter in each case compared to the standard case. In
case 2 we adopt a lower density wind with $A_*=0.4$ (corresponding to
$V_{\rm w}=2500\ {\rm km}\ {\rm s}^{-1}$ for the same value of
$\dot{M}_{\rm w} = 1\times 10^{-5}\ M_{\odot}\ {\rm yr}^{-1}$).  In case 3
we adopt, $E_{52}=10$; in case 4, $\varepsilon=0.5$; and in case 5,
$p=2.2$.  Finally, case 6 considers the ISM as the external medium with
$n_1=1$ and $\chi =1$, and all other parameters the same as in the standard
case.

In all cases, we start with $\gamma_b=100$, as appropriate for the afterglow
phase. We show results only for the initial period during which $\gamma_b
\ga 1/\Theta_b$, since our model does not apply to later times when the
lateral expansion of the bullet is important. Most of the initial kinetic
energy of the bullet is dissipated in the ambient gas during this early
period. In presenting our results we assume that the cosmological redshift
of the source, $Z_s\ll1$, because as it turns out, only nearby GRBs will
produce sufficient bow-shock flux to be detectable.

The numerical results for all six cases are shown in Figure 3.  We plot the
spectrum of the emission from the bow shock, with the vertical axis being
the luminosity per unit frequency ($L_{\nu}= 4\pi D^2 F_{\nu}$).  The light
curves appear to have similar features in all cases. At low frequencies,
$L_{\nu} \propto \nu^{2}$ because of synchrotron self-absorption, while at
high frequencies, $L_{\nu} \propto \nu^{-p/2}$ due to efficient electron
cooling.  The primary peak in each lightcurve is located at the synchrotron
self-absorption frequency (denoted as the peak frequency hereafter), while
the second peak is located at the cooling frequency. Case 2 has a faster
wind speed, hence a lower electron density than the standard case.
Therefore, case 2 has lower peak luminosities and higher cooling
frequencies. The observation time calculated in this case is also longer
because it takes more time for the bullet to decelerate. Case 3 has a
higher input energy than the standard case, but its peak luminosities are
almost the same as those of the standard case. Also it has lower peak
frequencies and higher cooling frequencies. This is caused by the fact that
the bullet has more kinetic energy, and so its deceleration requires more
time and a longer distance. As a result, the density of the wind is lower
than that of the standard case when the bow shock emission is calculated.
Case 4 assumes $\varepsilon = 0.5$, implying that more energy is deposited
into the bow shock by the bullet. Compared to the standard case, this case
has a higher peak luminosity and a lower cooling frequency. Case 5 assumes
a lower value for $p$. Compared to the standard case, it has a somewhat
lower peak luminosity and a higher peak frequency. Case 6 considers the ISM
as the ambient medium, implying a much lower electron density than the
standard case with a wind profile.  Consequently, the peak luminosity and
the peak frequency are lower, while the cooling frequency is higher. Also
note that in this case the peak luminosity increases with time, in contrast
to the standard case.

With these results at hand, we may now compare the thermal emission from
the main supernova shock with the non-thermal emission from the GRB
bow-shock in the optical regime. Type Ia supernovae (SNe Ia) have been used
as ``standard candles'' because they all have similar light curves and peak
absolute magnitudes. The absolute $B$-magnitude at maximum light ($M_B$) of
typical SNe Ia is $\sim -18.5$ (Vaughan et al. 1995), corresponding to
$L_{\nu} \simeq 1.3 \times 10^{28}\ {\rm ergs}\ {\rm s}^{-1} \ {\rm
Hz}^{-1}$ at $\nu = 6.8\times 10^{14}\ {\rm Hz}$. This is at least two
orders of magnitude more luminous than the emission from the GRB bow-shock
in our calculation. Type II supernovae can be generally divided into two
relatively distinct sub-classes, SNe II-L (``linear'') and SNe II-P
(``plateau''). The majority of SNe II-L have a nearly uniform peak absolute
magnitude, though there are a few exceptionally luminous SNe II-L (Young \&
Branch 1989; Gaskell 1992; Filippenko 1997). The average $M_B$ of the
majority of SNe II-L is $\sim -16.5$ (Gaskell 1992), corresponding to
$L_{\nu} \simeq 2 \times 10^{27}\ {\rm ergs}\ {\rm s}^{-1}\ {\rm Hz}^{-1}$.
This is still more than one order of magnitude brighter than the 
emission from the GRB bow-shock. 
But if the emission from the bow-shock persists
for a sufficiently long time (as in case 3 of our calculation), then it may
exceed the thermal SNe II-L emission due to the decline in the supernova
lightcurve.  SNe II-P show a very wide dispersion in the distribution of
their peak absolute magnitudes; i.e., $M_B$ spans the range from
-14 to -20 (Young \& Branch 1989), corresponding to 
$L_{\nu} \simeq 2 \times 10^{26}$ -- $5 \times 10^{28} 
\ {\rm ergs}\ {\rm s}^{-1}\ {\rm Hz}^{-1}$. Thus, the less luminous
SNe II-P (such as SN1987A), have peak luminosities in $B$-band which are
comparable to the emission from the GRB bow-shock. Of course, GRBs may
occur in a rare subset of supernovae that have very different optical
luminosities than the typical values mentioned above.

The bow shock emission is more easily detectable at either radio or X-ray
frequencies.  At a frequency of $\sim 5$ GHz, most radio supernovae which reached
their peak between 10 and 130 days had a peak spectral luminosity $L_{\nu}$
between $10^{26}$ and $10^{27}\ {\rm ergs}\ {\rm s}^{-1}\ {\rm
Hz}^{-1}$ (Li \& Chevalier 1999, Fig. 3). 
This peak luminosity is comparable to the bow shock emission in
Figure 3 (except for case 5 with $p=2.2$, for which $L_{\nu} \sim 10^{25}\
{\rm ergs}\ {\rm s}^{-1}\ {\rm Hz}^{-1}$). However, SN 1998bw, the most
luminous radio supernova observed so far, reached after $\sim 10~{\rm
days}$ a peak luminosity of $L_{\nu} \sim 10^{29}\ {\rm ergs}\ {\rm
s}^{-1}\ {\rm Hz}^{-1}$ at 5 GHz, which is about two order of magnitude
brighter than our calculated bow-shock emission.  

In the X-ray regime, the best studied case of Type II supernova SN 1993J,
had a luminosity of $3\times 10^{39}\ {\rm ergs}\ {\rm s}^{-1}$ in the
0.1-2.4 KeV band on day 7, corresponding to an average spectral luminosity
of $\sim 5.4 \times 10^{21} \ {\rm ergs}\ {\rm s}^{-1}\ {\rm Hz}^{-1}$ at
frequencies from $2.4\times 10^{16}$ to $5.8\times 10^{17} {\rm Hz}$.  The
X-ray flux from this supernova declined subsequently by $\sim 44 \% $ in
one month (Zimmermann et al. 1994; Fransson, Lundqvist \& Chevalier 1996).
These luminosities are an order of magnitude lower than the typical
bow-shock emission shown in Figure 3 (except for case 6 which considers the
ISM, where the two are comparable).  X-ray emission has been detected
within the first 100 days of a few other supernovae, such as SN 1980K, SN
1994I and SN 1998bw (see Table 1 in Pian 1999). SN 1980K had a luminosity of
$\sim 5\times 10^{38}\ {\rm ergs}\ {\rm s}^{-1}$ in the 0.2-4 KeV band on day 
44 and SN 1994I had a luminosity of $1.6\times 10^{38}\ {\rm ergs}\ {\rm
s}^{-1}$ in the 0.1-2.4 KeV band between day 79 and day 85.  Both of these
luminosities are much fainter than the bow-shock emission we calculated.
X-rays from the vicinity of SN 1998bw were first detected one day after the
explosion, and had a luminosity of $5\times 10^{40}\ {\rm ergs}\ {\rm
s}^{-1}$ in the 2-10 KeV band (Pian 1999), corresponding to an average
spectral luminosity of $\sim 2.4 \times 10^{22} \ {\rm ergs}\ {\rm s}^{-1}\
{\rm Hz}^{-1}$ at frequencies from $4.8\times 10^{17}$ to $2.4\times
10^{18} {\rm Hz}$.  This is comparable to the bow-shock emission shown in
Figure 3 (except for case 6 with the ISM), and so the X-ray emission from
SN 1998bw could have resulted from a bow-shock around a jet. However, the
association of the detected X-ray afterglow with SN1998bw is still not
certain due to the localization uncertainties (Pian 1999). In both the
radio and X-ray regimes, detection of the characteristic spectrum and time
evolution of the bow shock lightcurve can in principle be used to separate
it from the emission due to the main supernova shock.\footnote{Note that in
principle, there could also be a contribution from reflected X-rays due to
Compton scattering of the beamed GRB emission by the ambient gas (Madau,
Blandford, \& Rees 2000).  However, for the explosion parameters and photon
frequencies we consider here, this component appears to be negligible
compared to the bow shock emission. Moreover, its decay with time is much
faster than that of the bow shock component.}

Finally. we would like to find the maximum distances out to which the bow
shock emission would be detectable in the radio or X-ray regimes.  In the
radio regime, we adopt a detection threshold of 1~mJy at $10~{\rm GHz}$. In
the X-ray regime, we consider the sensitivity of the Chandra X-ray
Observatory (CXO), corresponding to a flux limit of $2\times 10^{-16}\ {\rm
ergs}\ {\rm cm}^{-2}\ {\rm s}^{-1}$ in the 0.5-2 KeV band [$=(1.2-4.8)\
\times 10^{17} {\rm Hz}$] for an integration time of 130 ks (Giacconi et
al. 2000). Obviously, this fiducial sensitivity can be improved with a
longer integration time.  The limiting distances corresponding to these
detection thresholds are listed in Table~1. We find that at $\nu = 10^{10}\
{\rm Hz}$ the bow shock emission can be detected out to $\sim 60$ Mpc when
the total energy input to the jets is equal to $5\times 10^{50}\ {\rm
ergs}$; in this case, the equivalent isotropic energy output of the GRB is
$10^{53}\ {\rm ergs}$ corresponding to case 3 in the calculation. In the
X-ray regime, detection is possible out to larger distances although there
might be a problem of confusion with other sources at the arcsecond angular
resolution of CXO.

\section{CONCLUSIONS}

Beamed GRBs produce a bow shock in their gaseous environment, which emits
non-thermal synchrotron radiation extending from radio to X-ray
frequencies.  We calculated the bow-shock luminosity from beamed GRBs in a
precursor wind or ISM environments (see Fig. 3), during the first few weeks
after the explosion.  For typical parameters, we find that the bow shock
emission from a jet with half-angle $\sim 5^\circ$ is visible out to tens
of Mpc in the radio and hundreds of Mpc in the X-rays (Table 1).  We
emphasize that the calculated bow shock luminosity is highly sensitive to
the total hydrodynamic energy carried by the jets and the density of the
ambient medium (see cases 2, 3 and 6 in Fig. 3).

The event rate for lines of sight outside the cones of the jet emission is
larger by a factor of $\sim 260 (\Theta_b/5^\circ)^{-2}$ than for lines of
sight which detect the $\gamma$-ray burst itself and are constrained to
pass through the jet. The rate of $\gamma$-ray events is estimated to be,
$\sim 2.5 \times 10^{-8}~{\rm yr^{-1}}$ per $L_\star$ galaxy for isotropic
emission (Wijers et al. 1998).  The Virgo cluster has a total luminosity of
$\sim 430L_\star$ (Woods \& Loeb 1999) corresponding to a total event rate
of $\sim 3\times 10^{-3} (\Theta_b/5^\circ)^{-2}~{\rm yr}^{-1}$ out to a
distance of $\sim 20~{\rm Mpc}$.  At larger distances, the mean density of
$L_\star$ galaxies (Folkes et al. 1999) implies that during an observing
period $\tau_{\rm obs}$ there should be at least one event at a distance of
$D\sim 90~{\rm Mpc} (\tau_{\rm obs}/10~{\rm yr})^{-1/3}
(\Theta_b/5^\circ)^{2/3}$. 
These estimates and Table 1 imply that future radio and X-ray observations
of SN remnants in the local universe can be used to test whether collimated
GRBs are associated with SNe.

\acknowledgments

This work was supported in part by grants from the Israel-US BSF
(BSF-9800343) and NSF (AST-9900877), and the NASA grant NAG5-7039.

\clearpage

\begin{table}
\begin{center}
\begin{tabular}{|c|c|c|c|c|c|}                              \hline
case & time 
& $L_{\nu}(\nu=10^{10}\ {\rm Hz})$ 
& $D_{\rm radio}$
& $L_{\nu}(\nu=1.2\times 10^{17}\ {\rm Hz})$ 
& $D_{\rm X-ray}$ \\
& (days) & $({\rm ergs}\ {\rm s}^{-1}\ {\rm Hz}^{-1})$ 
& $({\rm Mpc})$
& $({\rm ergs}\ {\rm s}^{-1}\ {\rm Hz}^{-1})$ 
& $({\rm Mpc})$ 
\\ \hline 
1 (standard) & 2.7 -- 13.2 & $2.2\times 10^{26}$ -- $5.7\times 10^{26}$ 
& 13.6 -- 21.8 & $2.4\times 10^{22}$ -- $9.4\times 10^{22}$ & 347 -- 687 \\ \hline 
2 ($A_\star = 0.4$) & 6.8 -- 32.9 & $2.9\times 10^{26}$ -- $6.5\times 10^{26}$ 
& 15.6 -- 23.3 & $6.8\times 10^{21}$ -- $2.6\times 10^{22}$ & 185 -- 361 \\ \hline 
3 ($E_{52} = 10$) & 27.4 -- 131.7 & $1.1\times 10^{27}$ -- $4.2\times 10^{27}$ 
& 30.3 -- 59.2 & $1.3\times 10^{22}$ - $5.2\times 10^{22}$ & 255 -- 511 \\ \hline 
4 ($\varepsilon = 0.5$) & 2.7 -- 13.2 & $2.9\times 10^{27}$ -- $3.5\times 10^{27}$ 
& 49.2 -- 54.1 & $1.1\times 10^{23}$ -- $4.6\times 10^{23}$ & 743 -- 1519 \\ \hline 
5 ($p=2.2$) & 2.7 -- 13.2 & $2.7\times 10^{25}$ -- $1.4\times 10^{26}$ 
& 4.8 -- 10.8 & $4.0\times 10^{22}$ -- $1.0\times 10^{23}$ & 448 -- 708 \\ \hline 
6 (ISM) & 36.3 -- 97.2 & $4.2\times 10^{25}$ -- $2.6\times 10^{26}$ 
& 5.9 -- 14.7 & $1.1\times 10^{21}$ -- $2.4\times 10^{21}$ & 74 -- 110 \\ \hline 
\end{tabular}
\end{center}
\begin{center}
\caption{Values of observed time, the luminosity at $\nu = 10^{10}\ {\rm
Hz}$, the detection threshold distance at $\nu = 10^{10}\ {\rm Hz}$, the
luminosity at $\nu = 1.2 \times 10^{17}\ {\rm Hz}$, and detection threshold
distance at $\nu = 1.2\times 10^{17}\ {\rm Hz}$.}
\end{center}
\label{table:1}
\end{table}

\begin{figure} [t]
\centerline{\epsfysize = 4.0in \epsffile{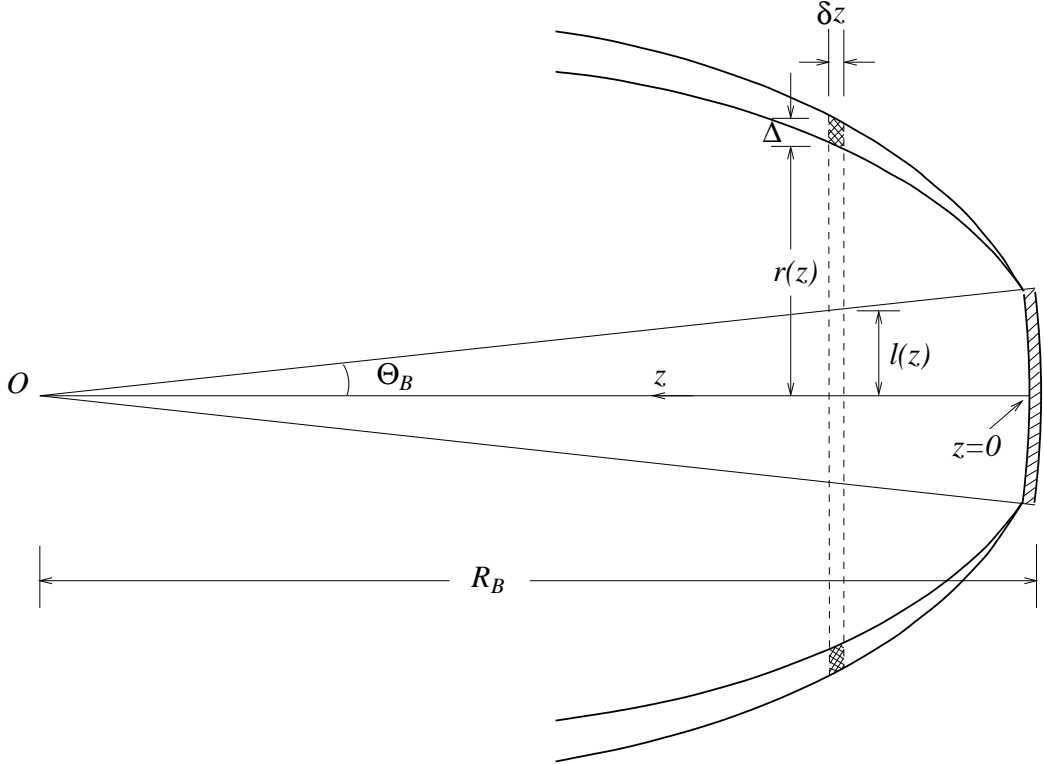}}
\caption{Geometry of the bow shock.}
\end{figure}

\begin{figure} [t]
\centerline{\epsfysize = 1.5in \epsffile{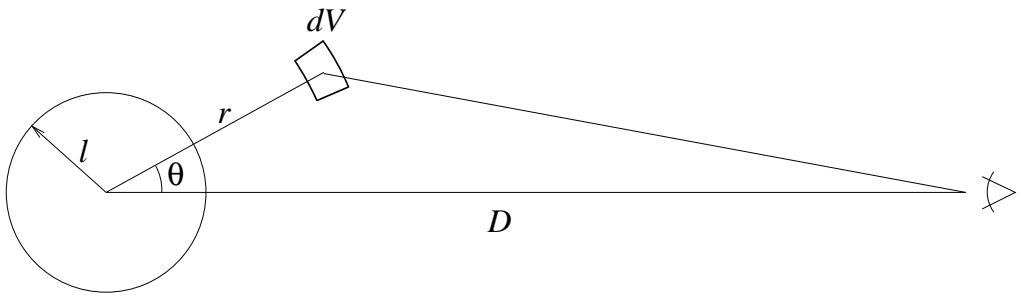}}
\caption{Coordinate system for the calculation of the bow shock emission.}
\end{figure}

\clearpage
\begin{figure} [t]
\includegraphics{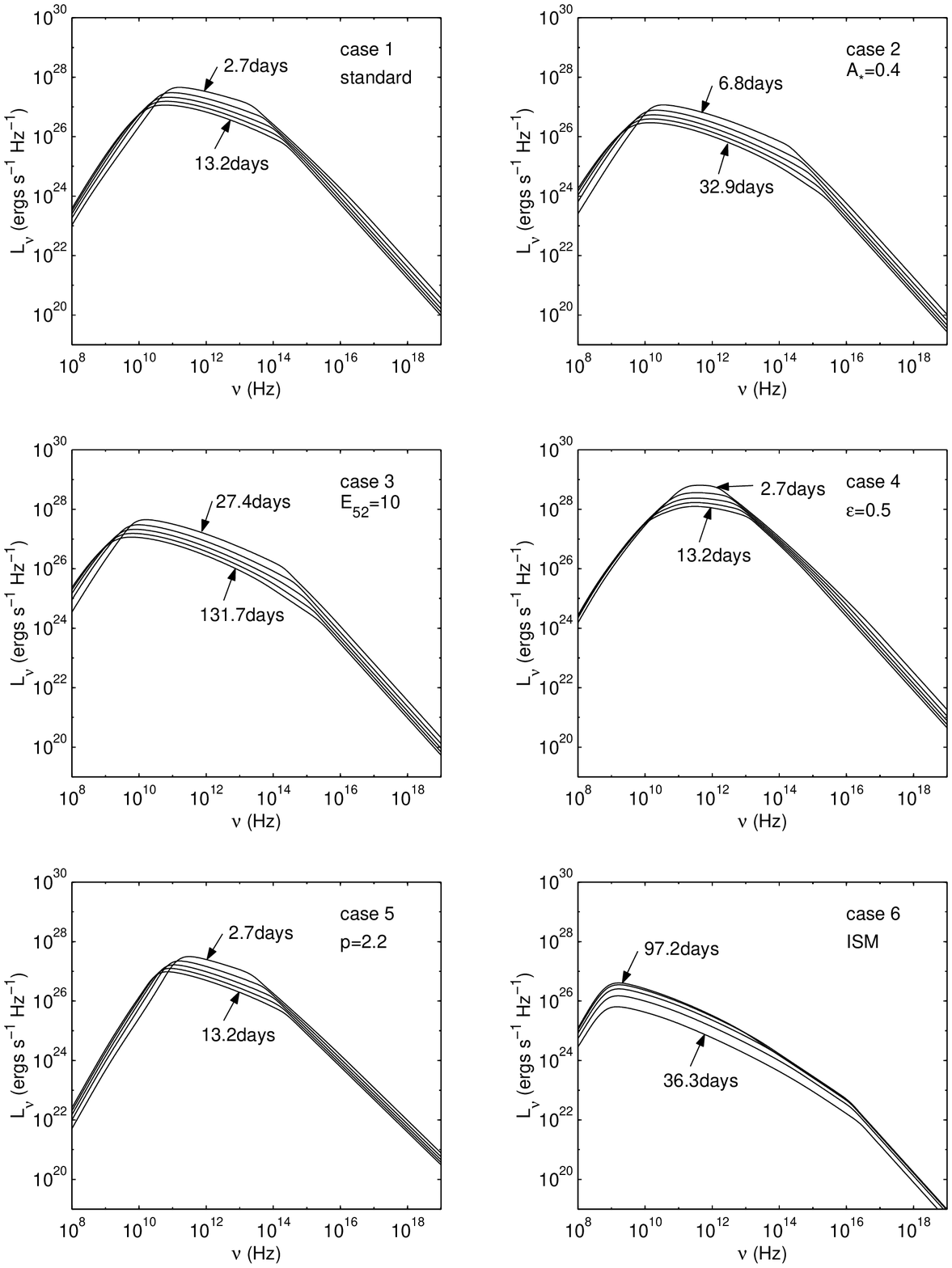}
\vspace{5.4in}
\caption{Lightcurves of the bow shock emission for six different cases (see
\S3 for details). In each case, the start and end times are labeled on the
plot. The intermediate times are evenly distributed between the start and
end times.}
\end{figure}

\begin{references}

\reference{} Berger, E., et al. 2000, Astro-ph/0005465

\reference{} Blandford, R. D., \& McKee, C. F. 1976, Phys. Fluids, 19, 1130

\reference{} Bloom, J. S. et al. 1999, Nature, 401, 453


\reference{} Chevalier, R. A. 1998, ApJ, 499, 810

\reference{} ---------------. 1999, ApJ, 511, 798

\reference{} Chevalier, R. A., \& Li, Z. Y. 1999, ApJ, 520, L29

\reference{} -----------------------------. 2000, ApJ, 536, 195

\reference{} Costa, E., et al.~1997, Nature, 387, 783

\reference{} Dai, Z. G., Huang, Y. F. \& Lu, T. 1999, ApJ, 520, 634

\reference{} Filippenko, A. V. 1997, ARA\&A, 35, 309

\reference{} Folkes, S., et al. 1999, MNRAS, 308, 459

\reference{} Frail, D. A., Kulkarni, S. R., Nicastro, L., Feroci, M., \&
Taylor, G. B. 1997, Nature, 389, 261

\reference{} Fransson, C., Lundqvist, P., \& Chevalier, R. A. 1996, ApJ, 461, 993

\reference{} Galama, T. J., et al. 1999, Astr. \& Ap.  S, 138, 465

\reference{} Garnavich, P. M., et al. 2000, ApJ, accepted (Astro-ph/0003429)

\reference{} Gaskell, C. M. 1992, ApJ, 389, L17

\reference{} Giacconi, R., et al. 2000, astro-ph/0007240

\reference{} Granot, J., Piran, T., \& Sari, R. 1999, ApJ, 513, 679

\reference{} Halpern, J. P., et al. 2000, Astro-ph/0006206

\reference{} Harrison, F. A., et al. 1999, ApJ, 523, L121

\reference{} Jensen, B. L., et al. 2000, Astro-ph/0005609

\reference{} Katz, J. I. 1994, ApJ, 422, 248

\reference{} Khokhlov, A. M., Hoeflich, P. A., Oran, E. S., Wheeler, J. C.,
Wang, L. 1999, ApJ, 524, L107

\reference{} Koyama, K., et al. 1995, Nature 378, 255 

\reference{} Koyama, K., et al. 1997, PSAJ, 49, L7

\reference{} Kulkarni, S. R., et al. 2000, to appear in Proc. of the 5th
Huntsville Gamma-Ray Burst Symposium; astro-ph/0002168


\reference{} Li, Z. Y., \& Chevalier, R. A. 1999, ApJ, 526, 716

\reference{} MacFadyen, A. I., Woosley, S. E., Heger, A. 1999, ApJ, submitted
(astro-ph/9910034)

\reference{} Madau, P., Blandford, R. D., \& Rees, M. J. 2000,
ApJ, in press; astro-ph/9912276

\reference{} Masetti, N., et al. 2000, Astro-ph/0004186

\reference{} M\'{e}sz\'{a}ros, P., \& Rees, M. J. 1993, ApJ, 405, 278

\reference{} -----------------------------------. 1997, ApJ, 476, 232

\reference{} Metzger, M. R., et al.~1997, Nature, 387, 879

\reference{} Muraishi, H. et al. 2000, A\&A, in press; astro-ph/0001047.

\reference{} Paczy\'{n}ski, B., \& Rhoads, J. E. 1993, ApJ, 418, L5

\reference{} Panaitescu, A., \& M\'{e}sz\'{a}ros, P. 1999, ApJ, 526, 707

\reference{} Pian, E. 1999, Astro-ph/9910236

\reference{}  Reichart, D. E. 1999, ApJ, 521, L111

\reference{} Rhoads, J. E. 1997, ApJ, 487, L1

\reference{} ------------. 1999a, ApJ, 525, 737

\reference{} ------------. 1999b, Astron. \& Ap. Suppl., 138, 539

\reference{} Rybicki, G. B., \& Lightman, A. P. 1979, 
Radiative Processes in Astrophysics (New York: Wiley Interscience), p. 147

\reference{} Sagar, R., Mohan, V., Pandey, S. B., Pandey, A. K., Stalin, C. S., 
\& Castro-Tirado, A. J. 2000, Astro-ph/0004223

\reference{} Sari, R., Piran, T., \& Narayan, R. 1998, ApJ, 497, L17

\reference{} Stanek, K. Z., Garnavich, P. M., Kaluzny, J., Pych, W., Thompson, I. 
1999, ApJ, 522, L39

\reference{} Tanimori, T. et al. 1998, ApJ, 497, L25

\reference{} Umeda, H. 2000, ApJ, 528, L89

\reference{} van Paradijs, J., et al. 1997, Nature, 386, 686

\reference{} Vaughan, T. E., Branch, D., Miller, D. L., \& Perlmutter, S. 
1995, ApJ, 439, 558

\reference{} Wang, X., \& Loeb, A. 2000, ApJ, 535, 788

\reference{} Waxman, E. 1997a, ApJ, 485, L5

\reference{} ---------. 1997b, ApJ, 489, L33

\reference{} Wijers, R. A. M. J., Rees, M. J., \& M\'{e}sz\'{a}ros, P. 
1997, MNRAS, 288, L51

\reference{} Wijers, R. A. M. J., Bloom, J. S., Bagla, J. S., \& Natarajan,
P. 1998, MNRAS, 294, L13

\reference{} Woods, E., \& Loeb, A. 1999, ApJ, submitted; astro-ph/9907110

\reference{} Young, T. R., \& Branch, D. 1989, ApJ, 342, L79

\reference{} Zimmermann, H. U., et al. 1994, Nature, 367, 621


\end{references}
\end{document}